\RequirePackage{fix-cm} 
\documentclass[pre,showpacs,
twocolumn,twoside,
amsmath,amssymb,amsfonts]{revtex4}
\usepackage{times}\usepackage{mathptmx}

\usepackage{graphicx}

\usepackage{microtype}
\usepackage{amsfonts}
\usepackage{mathtools}\mathtoolsset{centercolon}

\renewcommand{\d}{\mathrm{d}}

\newcommand{\name}[1]{#1}
\usepackage{url}
\usepackage[breaklinks,colorlinks]{hyperref}
\usepackage{breakurl}

\begin{document}

\title{Diffusion-limited reactions on a two-dimensional lattice with
  binary disorder}
\author{Andrea Wolff}\email{awolff@thp.uni-koeln.de}
\author{Ingo Lohmar}\email{il@thp.uni-koeln.de}
\altaffiliation[now at: ]{Racah Institute of Physics, The Hebrew University,
  Jerusalem 91904, Israel}
\author{Joachim Krug}\email{krug@thp.uni-koeln.de}
\affiliation{Institute for Theoretical Physics, University of Cologne,
  Z\"ulpicher Strasse 77, 50937 K\"oln, Germany}
\author{Yechiel Frank}\email{yechiel.frank@mail.huji.ac.il}
\author{Ofer Biham}\email{biham@phys.huji.ac.il}
\affiliation{Racah Institute of Physics, The Hebrew University,
  Jerusalem 91904, Israel}

\date{\today}

\begin{abstract}
  Reaction-diffusion systems where transition rates exhibit quenched
  disorder are common in physical and chemical systems.
  We study pair reactions on a periodic two-dimensional lattice,
  including continuous
  deposition and spontaneous desorption of particles.
  Hopping and desorption are taken to be thermally activated processes.
  The activation energies are drawn from a binary distribution of well
  depths, corresponding to `shallow' and `deep' sites.
  This is the simplest
  non-trivial distribution, which we use to examine and explain
  fundamental features of the system.
  We simulate the system using kinetic Monte Carlo methods and provide a
  thorough understanding of our findings.
  We show that the combination of shallow and deep sites broadens
  the temperature window in which the reaction is efficient, compared to
  either homogeneous system.
  We also examine the role of spatial
  correlations, including systems where one type of site is arranged
  in a cluster or a sublattice. Finally, we show that a simple rate
  equation model reproduces
  simulation results with very good accuracy.
\end{abstract}


\maketitle

\section{Introduction}
\label{sec:intro}

Reaction-diffusion systems are successful models to describe a large
variety of phenomena in physics, chemistry, and
biology~\citep{benavraham00,grzybowski05}.
They may involve one or more reactant species that diffuse and react
with each other on a surface or in the bulk.
In particular, surfaces often catalyze chemical reactions between
adsorbed atoms and molecules.
The densities of the adsorbed chemical species and their reaction rates
depend on parameters of the surface and on the temperature.
Microscopically, one can describe the diffusion of particles on the
surface as a random walk between adsorption sites.
In homogeneous systems all the adsorption sites are identical.
However, most systems are heterogeneous, involving different types of
adsorption sites with a broad distribution of binding energies.

The present study is motivated by a specific example of an
important surface process, namely the formation of molecular hydrogen on
dust grains in the interstellar medium~\citep{gould63, hollenbach70,
  hollenbach71a, hollenbach71b, smoluchowski83, duley86}.  Hydrogen
atoms impinge from the gas phase onto a grain, 
and diffuse on its
surface.  They may either desorb thermally from the surface, or
encounter each other and form a molecule.  This defines a
reaction-diffusion system in a spatially confined region.
In this article we are concerned with steady-state systems, when
the hydrogen
\emph{recombination efficiency} is defined as the fraction of impinging
particles that end up (and eventually desorb) in molecular form.  
This efficiency plays an important role in the evolution of interstellar
clouds.
Typically, there is a narrow window of temperatures in which
recombination is efficient.
At lower temperatures, the atoms are not sufficiently mobile to react,
whereas at higher temperatures they desorb too quickly. 

Assuming that all rates are spatially homogeneous, the system
is well understood analytically.  
A zero-dimensional master equation for
the particle number distribution~\citep{green01, biham02}, together with
a proper definition and calculation of the reaction rate coefficient in
terms of a first-passage problem~\citep{lohmar06, lohmar08}, suffices to
accurately describe the many-particle system~\citep{lohmar09a}.  
It is very important,
however, to consider disorder in the local rates of hopping and 
desorption 
of the particles.  As we alluded to earlier, this is not only of
theoretical interest.  In fact, in the astrophysical context, the
disordered case is much more realistic, and it is long known that
disorder potentially enhances the efficiency
dramatically~\citep{hollenbach71a}.
However, the combination of a
confined two-dimensional region, rate disorder and the many-particle
reaction-diffusion dynamics makes this problem notoriously hard to
tackle analytically.  
Kinetic Monte Carlo (KMC) methods can be used to simulate such
systems~\citep[e.g.][]{chang05, cuppen05}, and algorithms are still
subject to improvement~\citep[which also compares related
approaches]{tsvetkov09}.  They remain computationally expensive,
however, and a \emph{systematic} understanding of the effects of
disorder is still missing.

Here we start such an analysis for the simplest form of rate disorder,
where each lattice position corresponds to either a
standard (`shallow') site, or to a strong-binding (`deep') site,
with enhanced binding energy.
While we strive to keep this a
theoretical self-contained work, our models and questions are motivated
by applications and should easily translate to practice.  This is one
reason why we have chosen 
thermally activated rates throughout, and present
most results in terms of temperature, and on scales relevant to the
astrophysical problem just described.
In the latter context, our work is relevant to systems combining
physisorption (shallow sites) and chemisorption (deep
sites)~\citep{mennella08}
, aside from features particular to specific material systems.
Using such a discrete distribution turns out to be conceptually
different from the case of continuous distributions of binding
strength~\citep{chang05}, in which well depths drawn from tails of the
distribution may significantly affect the temperature window of high
efficiency.

Our goal in this paper is to provide a thorough understanding of all
relevant mechanisms 
of the described
reaction-diffusion system.  Most importantly, if we start from
homogeneous systems of either standard or deep sites, their temperature
windows of high efficiency will typically be separated by a gap.
It is a natural
question whether a mixture of the two types of sites still exhibits
two separated peaks, or whether (and under what conditions) the
efficiency is high for in-between temperatures.

Our findings are relevant for other systems as diverse as
catalysts~\citep{koper99}, exciton trapping in
photosynthesis~\citep{montroll69}, exciton transport in semiconducting
nanosystems~\citep{barzykin07}, and diffusion-limited reactions on
biomembranes~\citep{straube07}.  The generalization of our results to
these and other related contexts should be straightforward.

The paper is organized as follows.  In Sec.~\ref{sec:model} we define
the system and our notation.  The following Sec.~\ref{sec:qualitative}
provides a qualitative picture which identifies three temperature
regimes and describes the relevant processes in each.  In
Sec.~\ref{sec:kmc} we give a systematic account of extensive KMC
simulations and discuss the observed behavior in detail.  This includes
the study of spatial correlations in the quenched disorder.
Section~\ref{sec:re} presents a simple
yet accurate rate equation model, and we explain the difference to the
homogeneous case.  We derive an expression for the efficiency in the
most interesting regime.  Finally, we present our conclusions in
Sec.~\ref{sec:conclusion}.

\section{Model and Definitions}
\label{sec:model}

We consider a system of a single particle species on a two-dimensional
square lattice of $S$ sites with periodic boundary conditions.  Each
lattice site is characterized by a binding energy, which can take one of
two values --- we call this a \emph{binary lattice}.  The number of
sites of either type is denoted by $S_i$ ($i=1,\,2$), and $S=S_1+S_2$.
Particles impinge onto the lattice at a homogeneous rate $f$ per site.
If a site is already occupied, the impinging particle is rejected.  In
the context of surface chemistry this is known as
\name{Langmuir-Hinshelwood} (LH) rejection~\citep{langmuir18}.

Particles explore the lattice by hopping to neighboring sites with an
(undirected) rate $a$, and they can desorb from a site with rate $W$.
Both rates depend on the binding energy at the particle position.  If
two particles meet on one site, they form a dimer and leave the system
immediately.  The key quantity of such a system is the \emph{efficiency}
$\eta$, defined as the ratio between the number of particles that react and
the total number of impinging particles, when the system is in a steady
state.

In view of possible applications, we choose rates to be thermally
activated by a system temperature $T$.  The activation energy for
desorption is denoted $E_{W_i}$, which we identify with the binding
energy at the particle position.  Similarly, hopping from a type-$i$
site has an activation energy $E_{a_i}$.  All rates share the attempt
frequency $\nu$, so that, e.g.,
$W_i=\nu\exp\left(-E_{W_i}/T\right)$ --- here and in the following
energies are measured in temperature units.  We want to ensure
\emph{detailed balance}.  The simplest way to achieve this
 is by choosing $W_1/a_1=W_2/a_2$, or
equivalently, $E_{W_1}-E_{a_1}=E_{W_2}-E_{a_2}$, and we will employ this
choice throughout.  The number of sites visited by a single particle
before desorption 
becomes then independent of disorder.

\begin{figure}
  \centering
  \includegraphics{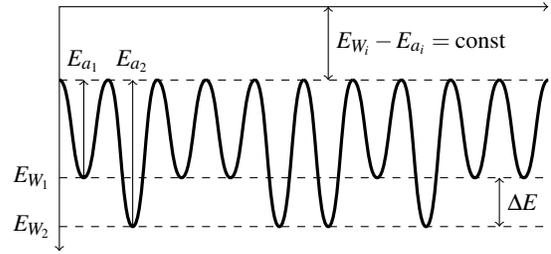}
  \caption{One-dimensional cut through the energy landscape of our model.}
  \label{fig:random-traps}
\end{figure}

To establish a connection to surface chemistry problems, we think of
type-$1$ sites as standard or `shallow' adsorption sites, and of
type-$2$ sites as strong-binding or `deep' sites, with
$E_{W_2}>E_{W_1}$.  A one-dimensional cut through such an energy
landscape is
sketched in Fig.~\ref{fig:random-traps}.

\section{Qualitative Discussion}
\label{sec:qualitative}

\subsection{Homogeneous systems}
\label{sec:qe-homogeneous}
For a homogeneous lattice, the dependence of the efficiency on the
temperature, $\eta(T)$, is known~\citep{green01,biham02}.  
At low
temperatures the particles are nearly immobile, thus they do not meet
other particles.  Therefore, the lattice is highly occupied, incoming
particles are mostly rejected, and the efficiency is low.  For higher
temperatures, hopping processes are activated and the particles begin to
explore the lattice.  This leads to more frequent encounters, so the
efficiency rises.  When the temperature is increased even further, the
particles tend to desorb before encountering each other, the lattice
coverage becomes small, and the efficiency decreases.

For the homogeneous system, the corresponding temperature bounds have
been obtained using rate equations~\citep{biham02}:
\begin{equation}
  \label{Tup}
  T^\mathrm{up}=\frac{2 E_W-E_a}{\ln(\nu/f)} 
\end{equation}
is the temperature
above which the kinetics becomes
second- (instead of first-) order, whence desorption ends the typical
particle residence and the efficiency is low, and 
\begin{equation}
  \label{Tlow}
  T^\mathrm{low}=\frac{E_a}{\ln(\nu/f)} 
\end{equation}
is the temperature
below which particles arrive faster than they hop, leading to
dominant LH rejection and low efficiency.  The average of these two
bounds reads 
\begin{equation}
  \label{Tmax}
  T^\mathrm{max}=\frac{E_W}{\ln(\nu/f)} 
\end{equation}
and corresponds to
the temperature of maximum efficiency.

If the binding energy $E_W$ is increased, the efficiency maximum is
shifted towards higher temperatures, and the shift is directly
proportional to the change in binding energy.  For a binding energy
difference $\Delta E=E_{W_2}-E_{W_1}$ of two (otherwise equal) lattices
of either type-$1$ or type-$2$ sites, the relation between the
temperatures of maximal efficiency is given by
\begin{equation}
 T^{\mathrm{max}}_{2} 
=T^{\mathrm{max}}_{1}\cdot\left(1+\frac{\Delta E}{E_{W_1}}\right),
\end{equation}
whereas the peak width $T^{\mathrm{up}}_i-T^{\mathrm{low}}_i$ is the same.

\subsection{Binary systems}
\label{sec:qe-binary}

Now we consider the binary lattice introduced in Sec.~\ref{sec:model},
with binding energies
$E_{W_1}$ and $E_{W_2}$.  To each site, we randomly assign a binding
energy.  There is a
typical length for a particle to find a strong-binding site.  This
length obviously shortens when there are more and more of these sites on
the lattice.  At low temperatures around the efficiency maximum of the
type-$1$ sites, particles can only diffuse on and desorb from these
shallow sites, while particles landing on or hopping onto strong-binding
sites cannot leave by hopping or desorption, since the binding energy is
too high.  Recombinations either take place on the shallow sites, or by
hopping to an occupied neighboring strong-binding site.  For very high
temperatures around the efficiency maximum of the deep wells, the
particles diffuse on, desorb from and recombine on those, while on the
shallow sites, they desorb too quickly to allow any other processes.
But in the intermediate temperature regime --- right of the shallow
peak, left of the strong-binding peak --- something different happens.
Here, the temperature is too low for dynamics on deep wells, so
particles encountering a deep well are stuck.  On the other hand,
particles on shallow sites tend to desorb rather quickly, and thus do
not recombine on such sites.  But if they find a deep well before
desorbing, they are trapped until another adatom shares their fate and
they recombine.

The simple random walk with traps has been studied 
extensively \citep[e.g.][]{montroll69, evans85}.
To leading order, the average number
of steps a random walker performs before trapping is given by
\begin{equation}
\label{ntrap}
 \langle n\rangle\approx\frac{1}{\pi} \frac{1}{S_2}S\ln S,
\end{equation}
where $S_2$ is the number of deep wells and $S$ is the total number of
sites on the lattice.
This leads to a trapping length
\begin{equation}
  \label{ltrap}
  \ell_\mathrm{trap}=\sqrt{\langle n \rangle}.
\end{equation}
On the other hand, the typical radius of the area a walker explores on
standard sites before desorption is the random walk
length~\citep{lohmar08}
\begin{equation}
  \label{lrw}
  \ell_\mathrm{rw}= \sqrt{\frac{a_1}{W_1}}. 
\end{equation}
Trapping now competes with desorption from shallow sites; the former
only depends on the number of traps $S_2$, while the latter is a
function of temperature.  As long as the random walk length
$\ell_{\mathrm{rw}}$ is larger than the trapping length
$\ell_{\mathrm{trap}}$, the particles are --- on average --- trapped
before they can leave the lattice.  For a given number of traps $S_2$
this implies a high efficiency approximately up to the temperature
$T^\mathrm{eq}$ where both lengths become equal.  If this temperature
lies above the intermediate temperature range where both pure systems
have poor efficiency,
we can expect a high efficiency throughout, hence a
full `bridging' of the gap. Since the efficiency is high over this
whole temperature range then, we call this an efficiency \emph{plateau}.
We will calculate the value of the efficiency on such a plateau in a
rate equation model in Sec.~\ref{sec:plateau-re}.

Summing up, we can divide the temperature axis into three regions.  The
lowest temperatures where only particles on shallow sites are mobile, the
intermediate regime where the particles behave like random walkers on a
lattice with traps, and the high temperatures where particles become
mobile on strong-binding sites.  We now check this qualitative picture
with KMC simulations.

\section{Kinetic Monte Carlo Simulations}
\label{sec:kmc}

\subsection{Setup}
\label{sec:kmc-setup}

In order to test our predictions, we carried out extensive kinetic Monte
Carlo simulations.  The standard algorithm proceeds as follows
\citep[cf.][for a review]{voter07}.  We keep track of the full
microscopic dynamics of continuous-time random
walkers~\citep{montroll65} with standard exponential waiting time
distributions.  In each simulation step, the current system
configuration determines the list of possible elementary processes and
their rates.  By comparing a random number with the normalized partial
sums of these rates we find the process to execute next.  The simulation
time is then advanced according to the total sum of rates and the
configuration is updated.

For a given realization, we wait for the system to reach the steady
state before we measure the efficiency over $10^6$ impingements.  We
use a square lattice of $S=100\times 100$ sites.
We choose the other model parameters inspired by an exemplary system in
the astrophysical application, to show the relevance of our work in this
field, and since the corresponding system is known to exhibit
interesting kinetic regimes.
The flux of hydrogen atoms per unit surface area 
depends on gas density and temperature.  The flux
per surface site is given by the ratio between the flux
per unit area and the density of adsorption sites, hence it depends on
the surface morphology.
More precisely, the flux per site is given by $f = \rho v/(4s)$,
where $\rho$ is the density of hydrogen atoms in the gas phase,
$v$ is their average thermal velocity and
$s$ is the density of adsorption sites on the surface.
To obtain typical values we use $\rho=10\ \mathrm{cm}^{-3}$,
$v=1.45 \times 10^5\ \mathrm{cm}/\mathrm s$ 
(which corresponds to a gas temperature of $100\ \mathrm K$)
and $s = 5 \times 10^{13}\ \mathrm{cm}^{-2}$ which is the
measured density of adsorption sites on the amorphous carbon
sample studied in Ref.~\citenum{biham01}.
This results in a flux per site of $f=7.3\times10^{-9}\ \mathrm s^{-1}$.
For the attempt frequency we choose the standard value of $10^{12}\
\mathrm s^{-1}$ which is commonly used throughout surface science.
With each site we associate either the standard binding energy
$E_{W_1}=658\ \mathrm{K}$, as found for hydrogen atoms on amorphous
carbon~\citep{katz99}, or an enhanced energy $E_{W_2}= E_{W_1}+\Delta E$
with $\Delta E= 250$, $750$ or $1500\ \mathrm{K}$.  The activation
energy for hopping reads $E_{a_1}=511\ \mathrm{K}$ or $E_{a_2}=
E_{a_1}+\Delta E$, respectively.

In each case, we determine the efficiency as a function of the
temperature $T$, as well as of the relative frequency of strong-binding
sites $S_2/S$.  We do this for up to four different ways of distributing
the binding strengths.  For dynamics with nearest-neighbor hopping of
the particles, we either \emph{randomly assign} to each site a binding
energy with probabilities $p_1$ and $p_2=1-p_1$, respectively,
or we arrange the
strong-binding sites in a regular \emph{sublattice}, or we concentrate
all strong-binding sites in a single square \emph{cluster}.  In the case
of random assignment, $S_2$ is then binomially distributed with
parameter $p_2$.  To eliminate the fluctuations in $S_2$, we average the
efficiency over $20$ realizations.
In the following discussion, we can therefore identify $S_i/S$ with its
average $p_i$.
For
comparison with the rate equation model to be introduced in
Sec.~\ref{sec:re}, we also implement another kind of dynamics
(\emph{`longhop' case}), namely hopping from any site to any other site
of the lattice. This switches off any spatial correlations between the
lattice sites and thus is best suited for comparison with an effective
zero-dimensional model.

Binary disorder models very similar to random assignment and the
clustered case have been simulated before~\citep{chang05}.  The authors
were predominantly concerned with showing that such models can exhibit
efficient reaction over a broader range of temperatures than homogeneous
systems.  Here we extend these findings to a systematic picture for the
effect of the deep-site fraction $p_2$ and the energy gap $\Delta E$.
More importantly, we provide detailed explanations and analytic results
which explain all notable features of the simulation outcome in terms of
microscopic physical processes.

\subsection{Results}
\label{sec:kmc-results}

\begin{figure*}
  \centering
  \includegraphics{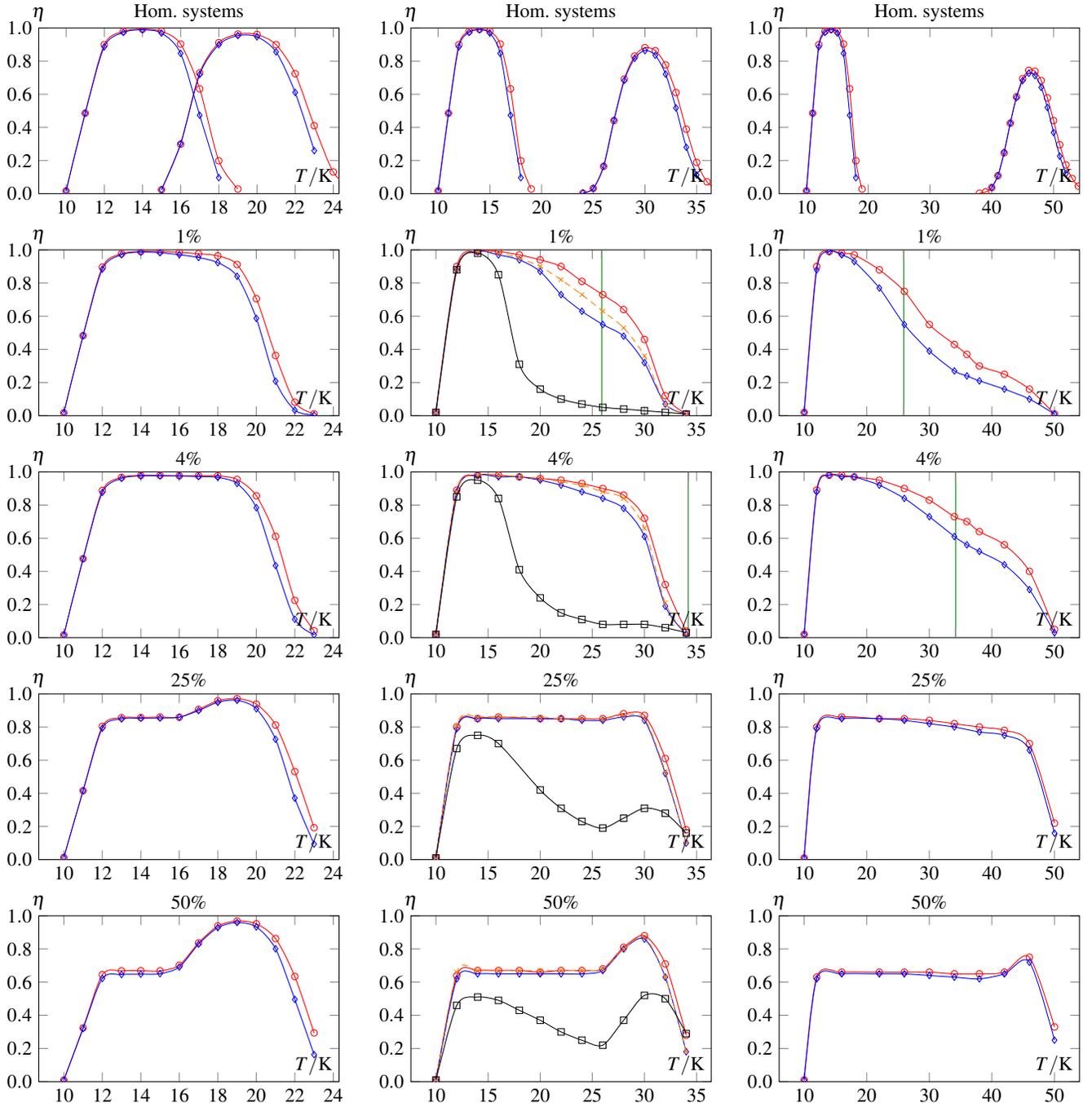}
  \caption{(Color online) Efficiency versus temperature for various
    fractions of deep sites.  Left column $\Delta E=250\ \mathrm K$,
    middle $\Delta E=750\ \mathrm K$, right $\Delta E=1500\ \mathrm K$.
    Randomly assigned energies (blue line, diamonds), longhop dynamics
    (red line, circles).  Only for $\Delta E=750\ \mathrm K$: sublattice
    (orange dashed line, crosses), and cluster (black line, squares).
    Vertical green line at $T^\mathrm{eq}$.  The first row shows the
    results for homogeneous systems of only standard or only deep sites,
    respectively.}\label{fig:eta-T-kmc}
\end{figure*}

Figure~\ref{fig:eta-T-kmc} shows the results of our simulations.  For
each $\Delta E$, we simulated systems with $1$, $4$, $25$ and $50\%$ of
strong-binding sites.  The random distribution is probably the most
interesting regarding applications.  Following the series of Figs.\ for
each $\Delta E$, we observe that the intermediate temperature regime is
bridged in each case.  This is in accordance with the analytic
prediction of Sec.~\ref{sec:qe-binary}, since already for moderate
deep-site fraction, the trapping length
$\ell_{\mathrm{trap}}$ is smaller than the random walk length
$\ell_{\mathrm{rw}}$ for all intermediate temperatures.
The observation
holds at least up to $\Delta E = 2500\ \mathrm K$ (not shown), which is
the largest value of $\Delta E$ that we have considered; beyond this
energy scale one enters the regime of chemisorption, which is not our
focus in this work.  The bigger the difference of the binding energies,
the more strong-binding sites are needed to form a genuine plateau,
where the efficiency does not depend on the temperature.  This complies
with the ideas of Sec.~\ref{sec:qualitative}; when the deep-site peak is
shifted to higher temperatures, $T^\mathrm{eq}$ has to increase to
warrant formation of a plateau.  This is achieved by increasing the
deep-site fraction.
The variance of the efficiency between different realizations of random
landscapes was found to be negligible throughout.  We also examined the
longhop case on such landscapes, and found that the efficiency varies
just as much.  Since this cannot be affected by any spatial
correlations, we conclude that this variation is always due to
fluctuations in the number of deep sites $S_2$ only.

\newlength{\fght}\setlength{\fght}{42.85mm}

The arrangement of strong-binding sites in a sublattice performs
slightly better, compared to random assignment.  This is not astonishing
since the sublattice optimizes the distance between the traps.  In the
random case, small clusters of strong-binding sites can occur, in which
a single trap is less efficient.  An alternative picture is that the
capture zones of individual traps typically have an overlap, which is
minimized in the sublattice case.

On a lattice with a single square cluster of strong-binding sites, there
is no bridging effect for any energy difference or frequency of
strong-binding sites.  For high frequencies of either shallow or
strong-binding sites, only one restricted peak emerges, while for
intermediate frequencies of strong and shallow sites two nearly
separated peaks appear.  
The efficiency does not drop to zero in the intermediate temperature
regime, because an exchange between shallow and deep sites takes place
along the boundary of the cluster.  However, since the boundary length
scales as $\sqrt{S}$, the fraction of boundary sites decreases with 
increasing $S$, and correspondingly the  
suppression of the efficiency in this regime becomes even more
pronounced for
larger systems.  We checked this for a system of $500\times500$ sites
(not shown).  This is in contrast to the well-mixed case, where a finite
fraction of sites are boundary sites (see below).

For the longhop case, we first verified that results on a sublattice and
a cluster landscape coincide, ensuring the correctness of the algorithm.
The efficiency for this kind of dynamics outperforms even the sublattice
results for nearest-neighbor hopping.  This is because in the
sublattice case, there is still the necessity for a particle to actually
travel to a trap instead of having a non-zero probability to reach a
trap on every step.  A further analysis of this model is provided in
Sec.~\ref{sec:re}.

\begin{figure}
  \centering
  \includegraphics{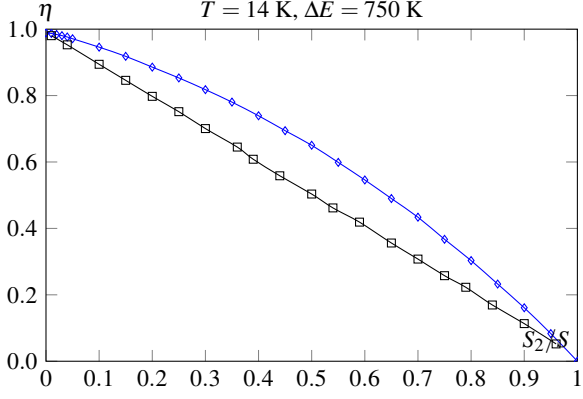}
  \caption{(Color online) Efficiency as function of the deep-site
    fraction for $T=14\ \mathrm K$ and $\Delta E=750\ \mathrm K$, for
    clustered deep sites (black, squares, $\eta_\mathrm{cluster}$) and
    randomly assigned energies (blue, diamonds,
    $\eta_\mathrm{random}$).}\label{fig:eta-s2-kmc}
\end{figure}

\begin{figure}
  \includegraphics{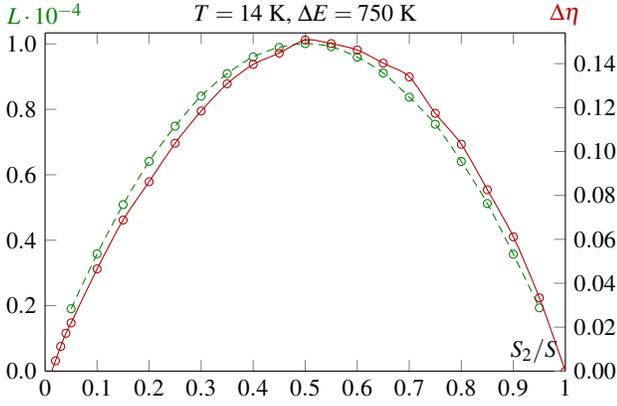}
  \caption{(Color online) Border length $L$ (green dashed line, left
    axis) and efficiency difference $\Delta\eta$ (dark red, right axis),
    as function of deep-site fraction, for $T=14\ \mathrm K$ and $\Delta
    E=750\ \mathrm K$.  Vertical axis scaling taken from data fit.}
  \label{fig:delta-eta}
\end{figure}

In addition to our qualitative explanations, we numerically examine the
dependence of the plateau efficiency value on the number of deep wells.
First we note that 
for our choice of parameters,
the efficiency value at $T^\mathrm{max}_1 \approx 14\ \mathrm{K}$
always corresponds to the plateau value.  
From the results for $\Delta E=750\ \mathrm{K}$ shown in
Fig.~\ref{fig:eta-s2-kmc} we infer that the way of distributing the
strong-binding sites is of crucial importance.  In the case of a single
square cluster of deep wells, the efficiency decreases linearly as
$1-S_2/S$, while for the random distribution the efficiency first
decreases more slowly (for less than $50\%$ of strong-binding sites) and
faster to the end (more than $50\%$).  We propose that 
this effect is related to the border length between shallow and deep sites,
and use this connection to derive an empirical formula for the plateau
efficiency.  For randomly distributed
deep wells, we calculate the border length $L$ as function of $S_2/S$
(cf.\ Fig.~\ref{fig:delta-eta}).  We find a shallow site next to a deep
site with probability $(S_2/S) (1-S_2/S)$.
Since the orientation of the pair does not matter, we gain an additional
factor of $2$.  Furthermore we have $2S$ possibilities to place such a
pair of sites on a square lattice with $S$ sites and periodic boundary
conditions.  So we find the following expression for the border length
between shallow and deep sites
\begin{equation}
 L= 4 S \cdot \frac{S_2}{S} \left(1-\frac{S_2}{S}\right).
\end{equation}
Fitting the efficiency difference $\Delta\eta =
\eta_{\mathrm{random}} - \eta_{\mathrm{cluster}}$ to a multiple of this
border length yields
\begin{equation}
 \Delta\eta=C \cdot L,
\end{equation}
with $C = (1.487 \pm 0.019) \times 10^{-5}$ or
\begin{equation}
  \label{etarandom}
  \eta_{\mathrm{random}}
  \approx \left(1-\frac{S_2}{S}\right)
  \cdot\left(1+(0.595\pm 0.008) \frac{S_2}{S}\right)
\end{equation}
for the empirical plateau efficiency value.
The quality of the fit $\Delta\eta\propto L$ for KMC results underlines
the role of the border length, and this corroborates our picture that
the dominant reaction process on the plateau is by hopping from standard
to deep sites. 
Further insight into the origin of Eq.~\eqref{etarandom} will be provided
below in Sec.~\ref{sec:plateau-re}.

\section{Rate Equation Model}
\label{sec:re}
Rate equations have been used previously to study reactions on finite
surfaces with different types of sites. Using surfaces with
varying roughness, where binding energies at a site are given by a
vertical bond strength plus an additional lateral bond strength per
in-layer neighbor of the landscape, KMC simulations were
performed~\citep{cuppen05}.  A rate equation model was then used to
check that such a rough landscape model is consistent with surfaces
deemed astrophysically relevant and examined in the laboratory.  To this
end, the rate equations with standard energy parameters were
time-integrated to predict the results of TPD experiments.

Here we apply a rate equation model to quantitatively reproduce our
KMC findings as well as to further our qualitative understanding of the
system's behavior.  From the definition of Sec.~\ref{sec:model} we
derive a set of rate equations for the total number $N_i$ of particles on
sites of type $i$.  Terms for desorption and influx are easily
written down, whereas the reaction terms are more subtle, partly since
in general, the reaction is not an elementary process with given rate.
For the time being, we denote the appropriate rate coefficients as
$A_i$, and refer to Sec.~\ref{sec:re-assumptions} for details.  The rate
equations then take the form~\citep{perets04,cuppen05}
\begin{equation}
  \label{rate-eqs}
  \begin{aligned}
    \frac{\d N_1}{\d t} &=
    f(S_1-N_1) -W_1N_1
    -A_1N_1(S_2-N_2) 
    -A_1N_1N_2 
    \\ &\quad
    -2A_1N_1^2
    +A_2N_2(S_1-N_1)
    -A_2N_1N_2, \\
    \frac{\d N_2}{\d t} &=
    f(S_2-N_2) -W_2N_2
    -A_2N_2(S_1-N_1) -A_2N_1N_2
    \\ &\quad
    -2A_2N_2^2
    +A_1N_1(S_2-N_2) -A_1N_1N_2.
  \end{aligned}
\end{equation}
Here the first two contributions cater for the impingement flux with
rejection and the desorption of particles.  For clarity we separated the
remaining terms.  The next two terms describe leaving to a site of
the opposite type (either to an empty or to an occupied site).  Then we
account for reactions inside one population due to hops
between sites of the same type, removing two atoms.  The remaining two
contributions describe gaining a particle by a hop from the other site
type, and finally, losing one particle due to the reaction with a
particle coming from the other population.

It is tempting to substitute the `internal' reaction term $2A_iN_i^2$ by
$2A_iN_i(N_i-1)$, since it should really depend on the number of
\emph{pairs}.  This is \emph{not} adequate: In the rate equation
treatment the $N_i$ are continuous and can drop below unity, such that
the reaction term (which we are ultimately interested in) could then
become negative.  The assumption that the reaction rate can be written
as above is at the heart of the rate equation approach (``mass action
law'').  Equations~\eqref{rate-eqs} are easily derived from the full
master equation using this assumption in the forms $\langle N_i(N_i-1)
\rangle \approx N_i^2$ and $\langle N_1N_2 \rangle \approx N_1N_2$
(where the expectation is over the joint probability distribution
$P(N_1,N_2)$ and the r.h.s.\ $N_i$'s are already the mean values as
above).

The reaction terms also provide the recombination rate of the process.
Adding up all terms proportional to the $A_i$ in $\d N/\d t =\d N_1/\d t
+\d N_2/\d t$, mere hopping terms (not leading to a reaction) cancel.
Using that the reaction consumes two particles, we obtain the rate at
which particles are removed by the reaction as
\begin{equation}
  \label{2R} 
  2R=2A_1N_1^2+2A_2N_2^2+2(A_1+A_2)N_1N_2,
\end{equation}
which can be simplified to $2R=2(A_1N_1+A_2N_2)(N_1+N_2)$.
Relating this to the particle influx $f(S_1+S_2)=fS$ gives the
\emph{efficiency} $\eta=2R/(fS)$.

\subsection{The reaction rate coefficient}
\label{sec:re-assumptions}

The homogeneous system was treated analytically by rate
equations~\citep{biham98, katz99, biham02}, the master
equation~\citep{green01, biham02}, and moment
equations~\citep{lipshtat03, barzel07b}.  For these methods just as for
stochastic or numerical methods based on these approaches, the reaction
rate coefficient is a crucial quantity, typically approximated as
$A\approx a/S$~\citep{stantcheva01}.  
We have argued elsewhere that this neglects the
nature of two-dimensional diffusion (``back diffusion'') as well as the
fundamental first-passage problem, the competition between a meeting
(hence reaction) of particles and the prior desorption of a reactant.
Hence we put some effort into a proper definition and evaluation of
$A$~\citep{lohmar06, lohmar08}, and we claimed that these
results should be applied in all mentioned frameworks, including the
rate equation treatment~\citep{lohmar09a}.

Here, we return to the choice $A_i=a_i/S$, since the situation is
different.
In rate equations such as
Eqs.~\eqref{rate-eqs}, there is no way to genuinely incorporate any
spatial structure.  This holds true for \emph{all} zero-dimensional
approaches, e.g., the master equation as well.  However, in the
homogeneous systems studied before, this neglect only concerns the
spatial correlations in the particle residence probability, with
well-studied effects~\citep{biham02, biham05, lohmar06}.  In the
heterogeneous system with its \emph{separated} populations, this
approach additionally neglects site type correlations.

For consistency, we are then forced to assume that a particle can reach any
other site by a single hop.
In particular, it hops to a site of type $i$ with probability $S_i/S$,
and it meets a particle on an $i$-site with probability $N_i/S$.
The conventional choice $A_i\approx a_i/S$ thus arises naturally if we use
rate equations to describe a system with site disorder, and we adopt
this choice in the following.
For a system with quenched spatial structure and nearest-neighbor hops
only, this description corresponds 
most closely to the well-mixed case.

\subsection{Comparison with KMC simulations}
\label{sec:re-kmc}
The rate equations~\eqref{rate-eqs} are exactly solvable at steady state by 
finding the real positive root of a third-order polynomial. 
However, the
results are cumbersome and less than illuminating.  We therefore
directly opted for a numerical solver throughout.

\begin{figure*}
  \centering
  \includegraphics{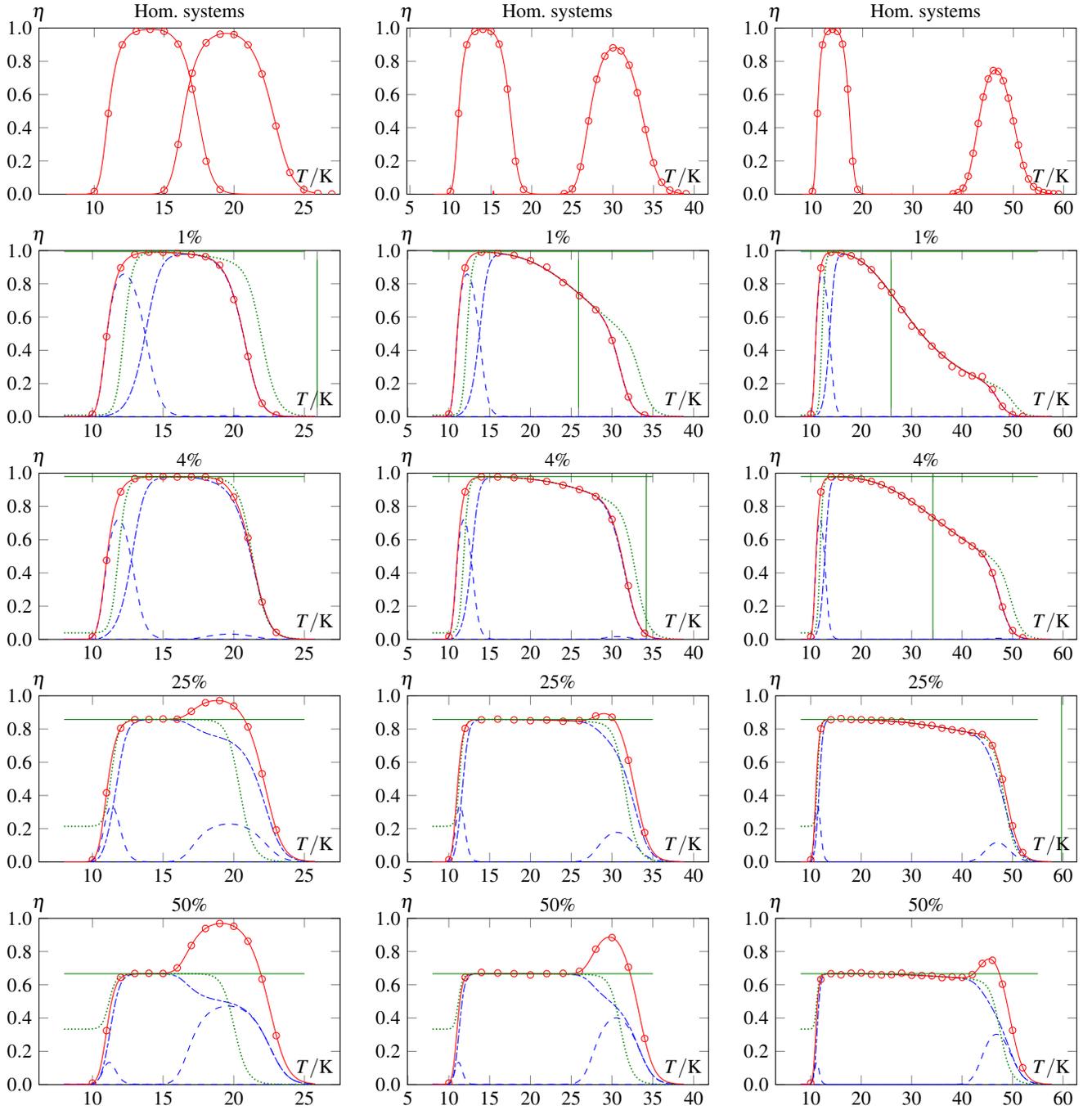}
  \caption{(Color online) Efficiency versus temperature for various
    fractions of deep sites.  Left column $\Delta E=250\ \mathrm K$,
    middle $\Delta E=750\ \mathrm K$, right $\Delta E=1500\ \mathrm K$.
    Red circles: KMC longhop results (see Sec.~\ref{sec:kmc}).  Red
    lines the numerical solution of rate equations with standard
    $A_i=a_i/S$ (solid), blue lines contributions by reaction on the
    $1$- and $2$-sites (dashed), and by switching between the types
    (dot-dashed).  Dotted green line the results of the plateau model,
    and green horizontal line the simple~\protect\eqref{plateau-eta}.
    Vertical green line at $T^\mathrm{eq}$.  The first row shows the
    results for homogeneous systems of only standard or only deep sites,
    respectively.}\label{fig:eta-T}
\end{figure*}

We find that the rate equations for the binary system
reproduce the outcome of extensive KMC (longhop) simulations for a
wide parameter range of practical relevance to excellent accuracy (see
Fig.~\ref{fig:eta-T}).
As noted in prior work~\citep{lohmar06}, however, since we present our
results as functions of temperature and parameters are thermally
activated, we typically have rather steep rises or declines, when even
factors of two or three in the efficiency need not appear substantial.
This hardly explains the overall accuracy, especially on plateaus and
moderate peaks for $\eta$ considerably smaller than unity.

Results on the validity of rate equations to describe the model in the
homogeneous case have shown that confinement to a finite surface renders
the discreteness of particles and fluctuations in the particle number
important~\citep{tielens95, biham02, krug03, biham05, lohmar06,
  lederhendler08}.  Consequently, the mean-field approach of rate
equations considerably overestimates the recombination efficiency in
small systems.  We do not see such effects for several reasons.
We are interested in the behavior of the system with a substantial
number of particles, when the effects of discreteness and of
fluctuations in this particle number are strongly reduced.  Further, the
confinement of particles to a finite surface is also far less important
than for the homogeneous system, because the majority of these particles
is trapped in deep wells in the regimes of most interest, anyway.
Finally, our system cannot be considered small, and we cannot preclude
completely that differences might be more pronounced for smaller system
sizes or different activation energies.

\subsection{Plateau efficiency}
\label{sec:plateau-re}
A key question of this work concerns the bridging between the two
efficiency peaks corresponding to homogeneous systems of one type of
sites.  
We have found a convincing qualitative picture before in
Sec.~\ref{sec:qualitative}, and we observe an efficiency plateau between
the two (virtual) peaks for a wide range of conditions.  What is the
value of the efficiency along this plateau?

We need to further simplify our rate equation model to arrive at a
simple analytic answer.
Several efforts to derive this from first principles 
were not met with success, hence we start 
from some observations:
Figure~\ref{fig:eta-T} shows that whenever a plateau emerges in
the efficiency, practically all recombinations are due to hops
between the two types of sites.
Only for very low concentrations of deep wells and when the efficiency
is close to unity, the lower temperature end of the plateau also includes a
substantial contribution from recombination on standard sites.
On the high-temperature end, any sizable contribution from reactions on
deep sites already results in an efficiency peak atop the plateau value
anyway.

For the plateau efficiency, we can therefore capture the essence of the
model accounting only for reactions \emph{between} the two populations.
Type $1$ denotes the standard sites, so we clearly have $A_1\gg A_2$.
We use this to neglect \emph{all} terms proportional to $A_2$, since
they are small compared to their $A_1$ counterparts, but 
keep all flux and desorption terms.  
Our reasoning is to retain as many terms
as possible, to remove those for recombination inside the $N_i$
populations, and since we can neglect the reaction $A_2N_1N_2\ll
A_1N_1N_2$, we have to leave out the corresponding $A_2$ hopping term
for consistency as well.
This leads to
the simplified steady-state equations
\begin{equation}
  \label{rate-eqs-simplified}
  \begin{aligned}
    0 &= f(S_1-N_1) -W_1N_1 -A_1N_1S_2, \\
    0 &= f(S_2-N_2) -W_2N_2 +A_1N_1S_2 -2A_1N_1N_2,
  \end{aligned}
\end{equation}
which yield an efficiency
\begin{equation}
  \label{eta-simplified}
    \eta_\mathrm{p}
    = \frac{2A_1N_1N_2}{fS} 
    = \frac{2fA_1S_1S_2(V_1+A_1S)}
    {S(V_1+A_1S_2)[V_2(V_1+A_1S_2)+2fA_1S_1]},
\end{equation}
where $V_i=W_i+f$.
We could now evaluate this at a
temperature right on the plateau.  It will turn out, however, that we
can make two more assumptions for this case.

First, we also neglect desorption from the $2$-sites, so $V_2=f$, and
using $A_1=a_1/S$, Eq.~\eqref{eta-simplified} reduces to
\begin{equation}
  \eta_\mathrm{p} = \frac{2(S_1/S)(S_2/S)(1+V_1/a_1)}
  {(V_1/a_1+S_2/S) (1+V_1/a_1+S_1/S)}.
\end{equation}
Second, on the plateau and for a reasonable deep-site fraction $S_2/S$,
we have $V_1/a_1\ll S_2/S <1$.  This yields
\begin{equation}
  \label{plateau-eta}
  \eta_\mathrm{p} \approx \frac{2}{S/S_1+1},
\end{equation}
which no longer depends on \emph{any} energy scales, and which we find
to be in excellent agreement with both KMC and full rate equation
results (Fig.~\ref{fig:eta-T}): Whenever a plateau forms (i.e., if
$\Delta E$ is large enough to separate the homogeneous-system peaks, and
if there are enough deep wells if $\Delta E$ is fairly large), the above
expression is valid.

To check the validity of these approximations, we
recall from Sec.~\ref{sec:kmc-results} that the peak temperature
$T^\mathrm{max}_1$ for standard-site
parameters was found to always belong to the
plateau.  
It is large enough not to lie on the low-temperature rise to
the standard-site peak, yet minimal so as not to depend on the peak
separation governed by $\Delta E$.
At $T=T^\mathrm{max}_1$, $V_1=W_1+f=2f$ and $V_2 =W_2+f =f
[(f/\nu)^{\Delta E/E_{W_1}}+1]$.  Reasonably, $f/\nu\lll1$, while the
smallest interesting $\Delta E\sim E_{W_1}-E_{a_1}$, such that the ratio
$\Delta E/E_{W_1}$ is not excessively smaller than unity.  This
justifies the approximation $V_2\approx f$, immediately eliminating
$\Delta E$ from the game, as suggested by Fig.~\ref{fig:eta-T}.  We now
check the order of $V_1/a_1 =2f/a_1
=2(f/\nu)^{(E_{W_1}-E_{a_1})/E_{W_1}}$.  The exponent is about $0.22$
for amorphous carbon, and with the corresponding standard flux we have
$V_1/a_1\approx 6.3\times 10^{-5}$ (cf.\ Sec.~\ref{sec:kmc-setup}).  This is
negligible compared to any interesting deep-well fraction $S_2/S$, which
completes the argument for Eq.~\eqref{plateau-eta}.  (We checked that
this holds at least equally well for standard olivine
parameters~\citep{katz99}.)

Knowing what terms can be neglected, this result is also easily derived
from further simplified rate equations.  We rather provide an intuitive
explanation.
We consider the system in the steady state, so all particles entering
the system also have to leave.
They enter by impingement to any site, and leave only from $2$-sites,
by LH rejection or by reaction with an incoming $1$-particle.
Consequently, particles from $1$-sites arrive at a rate $fS_1/S_2$ at
each $2$-site.  This implies a rate $2fS_1/S_2\cdot N_2$ of particles to
leave the system, as the reaction takes away \emph{two} atoms.
Alternatively, particles leave by LH rejection (rate-wise, this is
merely a separate desorption process) at a rate $f\cdot N_2$.
The efficiency is the fraction of impinging particles that react; in the
steady state, this is just the rate at which particles leave due to
reaction, normalized by the total rate to leave (by reaction or by LH
rejection).  This yields
\begin{equation}
  \label{etap}
  \eta_\mathrm{p} = \frac{2S_1}{2S_1+S_2},
\end{equation}
which coincides with Eq.~\eqref{plateau-eta}.
We note that Eq.~\eqref{etap} can be rewritten as 
\begin{equation}
  \label{etap2}
  \eta_\mathrm{p} = \frac{1 -S_2/S}{1 - S_2/(2S)} \approx 
  \left( 1 - \frac{S_2}{S} \right) \left(1 + \frac{S_2}{2S} \right)
\end{equation}
for $S_2/S \ll 1$, which is precisely of the form of the empirical
relation~\eqref{etarandom}.  The coefficient inside the second bracket
in Eq.~\eqref{etarandom} deviates from $1/2$ because it was
obtained through a fit over the entire range of $S_2/S \in [0,1]$,
whereas Eq.~\eqref{etap2} is strictly valid only when $S_2/S$ is small.

\section{Conclusions}
\label{sec:conclusion}
We have studied diffusion-limited reactions 
of particles on a
two-dimensional lattice which consists of shallow and deep sites, using
KMC simulations and rate equations.
In the case when the two types of 
sites are randomly mixed, we found that
the temperature range in which the reaction is efficient dramatically
broadens compared to a homogeneous system that includes only shallow or
only deep sites.
The rate equations are found to provide a good description of the
system and are in perfect agreement with the KMC results.
We have also studied a system in which the deep sites are clustered
together.  In this case the hopping between shallow and deep sites is
suppressed.
As a result, the recombination efficiency is dramatically reduced in
comparison with the case in which the shallow and deep sites are
randomly mixed.

We expect that the qualitative features observed for the binary
distribution will hold for a broader class of models with different
distributions of binding energies.
The results presented in this paper are also relevant in the context of
molecular hydrogen formation in the interstellar medium.
More specifically, high abundances of molecular hydrogen are observed
in photon-dominated regions~\citep{habart04}.
In these regions, the grain temperatures are too high to form molecular
hydrogen from weakly adsorbed hydrogen atoms.  It was proposed that
strong-binding sites in conjunction with the weak-binding sites enable
the efficient formation of molecular hydrogen under these
conditions~\citep{cazaux02,cazaux04}.
Our work provides a quantitative basis for this mechanism.

\begin{acknowledgments}
  This work was supported by Deutsche Forschungsgemeinschaft within
  SFB/TR-12 \emph{Symmetries and Universality in Mesoscopic Systems} and
  the Bonn-Cologne Graduate School of Physics and Astronomy, and by the
  US-Israel Binational Science Foundation.  JK acknowledges the kind
  support and hospitality of the Hebrew University through the Lady
  Davis Fellowship Trust.
\end{acknowledgments}

\end{document}